\documentclass[aps,pre,twocolumn,epsfig,twoside,showpacs]{revtex4}
\usepackage{graphicx} 
\begin{document}

\title{Inherent Structures and Kauzmann Temperature of Confined
Liquids}

\author{A. Attili$^\dag$, P. Gallo$^{\dag,\ddag}$\footnote[1] 
{Author to whom correspondence  should be addressed; e-mail: 
gallop@fis.uniroma3.it}
and M. Rovere$^{\dag,\ddag}$}
\affiliation{
  $^\dag$ Dipartimento di Fisica, Universit\`a ``Roma Tre''\\
  $^\ddag$ INFM Roma Tre and Democritos National Simulation Center \\
Via della Vasca Navale 84, 00146 Roma, Italy.}


\begin{abstract}
Calculations of the thermodynamical properties of a supercooled liquid
confined in a matrix are performed with an inherent structure analysis.
The liquid entropy is computed by means of a thermodynamical integration
procedure. The contributions to the free energy of the liquid can be decoupled 
also in confinement in the configurational and the vibrational part.  
We show that the vibrational entropy can be
calculated in the harmonic approximation as in the bulk case.
The Kauzmann temperature of the confined system is estimated
from the behavior of the configurational entropy. 
   
\end{abstract}

\maketitle

\section{Introduction}

It is well known that most
liquids upon supercooling undergo a transition to an amorphous
state, where mechanical properties typical of a solid phase combine
with a microscopically disordered 
structure~\cite{pablo,angell1,stillinger,deben+still}. 
Just below the melting
temperature the supercooled liquids manifest a slowing down of
dynamics. This behavior has been successfully interpreted in terms
of the Mode Coupling Theory (MCT) which is able to predict
the asymptotic properties of the density correlators upon
decreasing temperature on approaching a
temperature $T_C$~\cite{goetze}. This temperature
marks a crossover from a region where the
exploration of the phase space of the system is
determined by structural relaxations, to a region where
it is determined by hopping processes.
In the last few years theoretical approaches based on the
analysis of the potential energy landscape (PEL) of the 
supercooled liquid 
have driven a significant progress in the study of the thermodynamics of
the glass transition
below $T_C$~\cite{deben+still,sastry+deben,sciortino,heuer,parisi1,%
cavagna,angelani,coluzzi,lanave,sri,speedy,keyes}. 

The phenomena related to the glass transition are yet not well
understood in the case of liquids in confined geometries
or at contact with solid surfaces although these situations are
very relevant for many technological and biological applications.
It is in fact still not clear how the theoretical approaches developed
for bulk supercooled liquids can be extended to describe the
corresponding phenomenology when liquids are confined.

While it has been shown that MCT works also for interpreting 
the dynamics of confined liquids in several 
cases~\cite{pellarin1,pellarin2,noiwater1,noiwater2,krakoviack},
no studies of dynamics in confinement below $T_C$ have
so far been performed. It is therefore very relevant  
to study how the PEL and the thermodynamical
properties below $T_C$ are modified by
the presence of confinement.
Recently a mean field analysis of the PEL for thin films
has shown that confinement could affect the thermodynamical behaviour and 
the glass transition~\cite{truskett04}. 

As proposed by different authors~\cite{goldstein,stil+web} 
the behavior of a bulk supercooled liquid is determined by 
the dynamics of the system in and between the basins of the PEL.  
At low enough temperature there are two separated regimes,
the dynamics on the short time
scale can be described as the motion around the local minima, while
the long time dynamics is related to the transition between
different basins of energy. This separation of regimes has been
framed by Stillinger and Weber (SW)~\cite{stil+web} in the formalism of
the inherent structures (IS). According
to their definition an IS is the configuration of local
minima of the PEL. A basin is the set of points which maps to the same
IS under a local energy minimization performed by a steepest descent
procedure starting from a configuration equilibrated
at a certain temperature. In the SW formulation
under the assumption that
the basins with the same IS energy $e_{IS}$ have equivalent properties
in the canonical partition function the motions between 
different basins and the vibrations inside a single basin can be decoupled. 
This formulation allows one to define 
and study a configurational entropy $S_{conf}$ . 
This quantity, which represents
the difference between the liquid and the disordered solid entropies,
plays a central role in understanding the glass 
transition.

In the process of cooling the configurational entropy decreases and
eventually vanishes at a finite temperature, defined as the
the Kauzmann temperature $T_K$~\cite{kauzmann}.
In the interpretation
of Adam, Gibbs and Di Marzio~\cite{dimarzio,gibbs,deben+still2}
at $T_K$ an ideal thermodynamical transition should take place from the
supercooled liquid to an amorphous phase with a single 
configuration~\cite{deben+still,deben+still2,deben+still3,%
parisi1,coluzzi,coluzzi1}. The singular behavior of thermodynamical
quantities measured in experiments
at the conventional glass transition temperature
$T_g$ would be related to the {\it true} transition occurring
at $T_K < T_g$. 

It is generally found that phase transitions in confined fluids
are modified by confinement both from geometric effects and
the interaction with the substrate. This is particularly true
when large spatial correlations are expected to take place
and finite size effects could influence the transition
in a fluid confined in a restricted environment. This would be
also the case for the glass transition at $T_K$ if it
is interpreted in terms of a second order
phase transition~\cite{berthier} or in the framework of the
mosaic state scenario~\cite{bouchaud}.

Here we consider the case of a glass forming confined
liquid, a Lennard-Jones
binary mixture (LJBM), embedded in a disordered array of soft spheres.
Molecular Dynamics simulations have been performed for this system
upon cooling  
and a numerical test of MCT properties has been carried out.
The mixture follows also in this confining environment, as in the bulk,
MCT predictions very well~\cite{pellarin1,pellarin2}. Nonetheless
important differences due to confinement are found.
In particular the range of validity of the MCT predictions suffers
a reduction of $60\%$ with respect to the bulk. 
We found a crossover temperature $T_C = 0.356$ (in Lennard-Jones units)
to be compared with
the bulk value $T_C=0.435$ (in Lennard-Jones units)~\cite{kob} 
and therefore we observed
a reduction of circa $20\%$ of $T_C$ in going from the bulk to the
confined LJBM. 

We performed in this paper an IS analysis to evaluate   
the IS distributions, the temperature dependence of the
configurational entropy and finally 
the Kauzmann temperature of the confined LJBM
to be compared with the corresponding values for the LJBM 
in the bulk phase~\cite{sciortino,coluzzi}.
The paper develops as follows: in the next section we report 
computational details. The third section is devoted to the 
calculation of the IS for the confined LJBM. In the fourth 
section we evaluate the configurational entropy and the Kauzmann 
temperature. The last section is devoted to the conclusions.

\section{Computer simulation of the confined liquid upon supercooling}

We studied the LJBM proposed in ref.~\cite{kob} embedded in a rigid 
disordered array of 16 soft spheres.
The liquid binary mixture is composed
by 800 particles of type A and 200 particles of type B. The parameters of 
the Lennard-Jones potential are 
$\epsilon_{AA}=1$, $\sigma_{AA}=1$, $\epsilon_{BB}=0.5$,
$\sigma_{BB}=0.88$, $\epsilon_{AB}=1.5$, and $\sigma_{AB}=0.8$.
In the following the LJ units will be used.
The A and B particles interact with the soft spheres with a 
potential $V(r)=\epsilon (\sigma/r)^{12}$ where   
$\epsilon_{SA}=0.32$, $\sigma_{SA}=3$, 
$\epsilon_{SB}=0.22$, $\sigma_{SB}=2.94$.

Molecular dynamics simulations have been performed in the NVT ensemble 
along a isochoric path at various temperatures
upon cooling. The box length is fixed to $L=12.6$.
In previous work we already investigated 
the system in the range of
temperature from $T=5.0$ to $T=0.37$ and
further MD simulation
details are reported in ref.~\cite{pellarin1,pellarin2}.

Starting from the equilibrated configurations at the following
temperatures: $T=5,2,0.8,0.6,0.55,0.5,0.475,45,0.425,0.38$, we
performed new simulations for each temperature in order to obtain a number of 
equally spaced configurations and calculate the corresponding
IS for each temperature. 

The IS have been obtained by the
conjugate-gradient minimization procedure described
in the literature and adapted to our confined system. 
For each temperature $1000$ configurations have been minimized.  
The Hessian matrix has been diagonalized at each IS 
to calculate the eigenfrequencies.

\section{Inherent structure analysis}

In the SW formulation of IS the canonical partition function 
can be written as follows:
\begin{eqnarray}
\lefteqn{Z_N(T) = } \nonumber \\ 
&{}{}& \int de_{IS} \Omega \left( e_{IS} \right) exp 
\{ -  [ e_{IS} + f(T,e_{IS}) ] / k_BT \} 
\label{eq:1}
\end{eqnarray}
where $\Omega(e_{IS})$ is the number of distinct basins with
energy $e_{IS}$ and $f(T,e_{IS})$ is the free energy of 
the system restricted to
a single basin with energy $e_{IS}$.
The configurational entropy $S_{conf}$ can be
defined as  $S_{conf}=k_B ln(\Omega(e_{IS}))$. 
The energies of the IS are distributed with a probability 
given by
\begin{eqnarray}
\label{eq:2}
\lefteqn {P \left( e_{IS},T \right) =} \nonumber \\
&{}{}& \frac { exp
[ - \left( e_{IS} + f \left(T,e_{IS} \right)-TS_{conf}
\left(e_{IS} \right) \right) 
 / k_BT ] }{
Z_N(T) }
\end{eqnarray}

The configurational entropy can be defined also as the 
difference between the liquid entropy
and the entropy of the disordered solid (DS)
\begin{equation}
\label{eq:3}
S_{conf} = S_{liquid} - S_{DS}
\end{equation}

\begin{figure}
\includegraphics[width=80 mm]{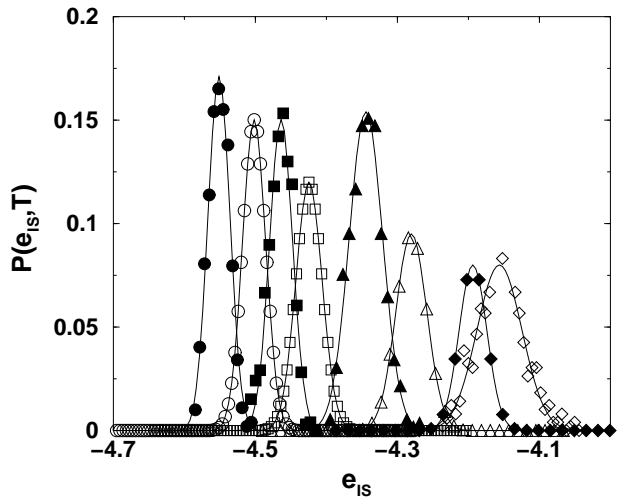}  
\caption{Distribution functions $P \left( e_{IS},T \right)$.
Lower temperatures are on the left. 
The functions are omitted for temperatures
$T=0.50$ and $T=0.55$ since they are almost overimposed to the ones
at $T=0.475$ and $T=0.60$ respectively.  
}
\label{fig:1}
\end{figure}

From the quenches performed at each temperature we
calculate the distribution functions $P \left( e_{IS},T \right)$
of the IS.
These are shown in Fig.~\ref{fig:1} together with
Gaussian best fits. The gaussian curves appear to reproduce
sufficiently well the distribution functions. For the lowest
temperatures the curves are narrower and more peaked around
the average value. 

\begin{figure}
\includegraphics[width=80 mm]{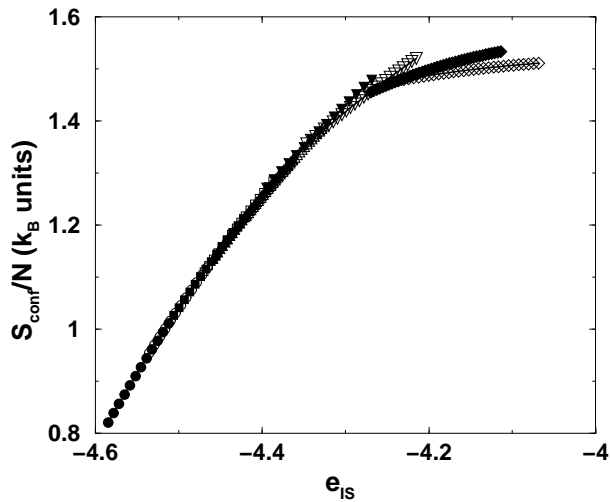}  
\caption{Entropy per particle calculated from the distribution functions of
the inherent structure energy according to Eq.~\ref{eq:6} 
at the different temperatures: $T=0.38,0.425,0.45,0.475,0.50,0.55,
0.60,0.80,2.0,5.0$.
The unknown temperature dependent term is
obtained by maximizing the overlap between the curves.
For $T>0.80$ the curves deviate from the
master curve.}
\label{fig:2}
\end{figure}

We can now look at the behavior of the configurational entropy
by considering Eq.~(\ref{eq:2}), from which we obtain 
\begin{eqnarray}
\label{eq:6}
\lefteqn{ln \left[ P \left( e_{IS},T \right) \right] + e_{IS}/k_BT = }
\nonumber \\
&{}{}& S_{conf}\left(e_{IS} \right)/k_B- f \left( T,e_{IS} \right)/ k_BT 
-ln \left[ Z_N(T) \right]  
\end{eqnarray} 
The left hand side of this equation can be calculated from 
the distribution functions of Fig.~\ref{fig:1} to obtain
a new set of curves.
By plotting 
all these curves as function of $e_{IS}$ we see that they
can be superimposed by subtracting a temperature dependent term
as shown in Fig.~\ref{fig:2} for $T < 0.80 $. Looking at the
right hand side of Eq.~\ref{eq:6}
this result implies that the the basin free energy $f(T,e_{IS})$ is almost
independent of the IS energy. 
The master 
curve represents, apart from an unknown temperature dependent term, the 
configurational entropy.    

Since the basin free energy is approximately
independent of $e_{IS}$ for $T<0.8$
in the confined liquid, the
partition function defined in Eq.~\ref{eq:1} can be separated as
\begin{eqnarray}
\lefteqn{Z_N(T) \approx } \nonumber \\
& exp [- f(T,e_{IS})  / k_BT ] 
\int de_{IS} \Omega \left( e_{IS} \right) exp 
\left( -  e_{IS} / k_BT \right). 
\label{eq:6b}
\end{eqnarray}

The confined liquid at low enough temperature can be
assumed to be composed of an inherent structure subsystem in thermal
equilibrium with the vibrational subsystem. The IS represents 
the long time dynamics of the system due to transitions between
the different basins of energy, whose degeneracy is counted
by $\Omega \left( e_{IS} \right)=exp(S_{conf}/k_BT)$. 
The vibrational spectrum is related to the oscillations close
to the minimum of the single basin. It
can be obtained by diagonalizing the Hessian matrix
after a conjugate gradient minimization starting from
equivalent state points~\cite{attili1}. 
 
\begin{figure}
\includegraphics[width=80 mm]{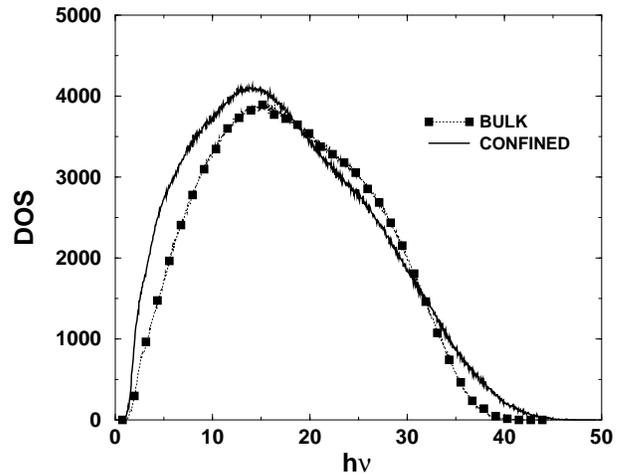}  
\caption{Density of state (DOS) of the confined liquid
compared with the DOS of the bulk liquid. The DOS are
obtained by quenching the systems from $T=0.38$ for the confined
mixture and from $T=0.51$ for the bulk. Both systems were
equilibrated at the same pressure before the quench,
the temperatures are different due to the fact that $T_C$ 
is lower for the confined system~\protect\cite{attili1}.
}
\label{fig:3}
\end{figure}

In Fig.~\ref{fig:3} we report the comparison of the density of state (DOS)
of the vibrational spectrum of the confined and the bulk LJBM
obtained with the same method.
We observe that the confinement does not induce large changes in 
both the shape and the spectral range of the eigenfrequencies. 
For the bulk LJBM it has been also shown that
the basin free energy can be approximated with the harmonic 
vibrational contribution~\cite{sciortino}. We will come back later
to this point.

\section{Configurational entropy and Kauzmann temperature}

The behavior of the configurational entropy has been determined 
from Eq.~\ref{eq:6} and shown in Fig.~\ref{fig:2} but the calculation
of the Kauzmann temperature requires the absolute value of $S_{conf}$.

A thermodynamical integration procedure 
allows one to evaluate the full entropy of the liquid
including the temperature dependent integration constant
which appears in Eq.~\ref{eq:6}. Starting from a
state reference point at temperature $T_r$ at the given 
volume $V$ of the simulation box,
the entropy $S_{tot}$ can be computed as
\begin{equation}
\label{eq:4}
S_{tot}(T,V) = S_{ref}(T_r,V) + \int_{T_r}^T \left( \frac 
{\partial U(T')}{\partial T'} \right)_V dT'
\end{equation}
where $U(T)$ is the internal
energy calculated in the simulation along the isochoric path.
As reference point we assume $T_r=5.0$.
The reference entropy $S_{ref}(T_r,V)$ is derived from the 
corresponding $S_{bulk}(T_r,V)$ of the bulk system at the same
temperature $T_r$ and volume $V$ by adding the contribution of the
work needed to include the soft spheres by keeping the
volume constant.
\begin{eqnarray}
\lefteqn{S_{ref}(T_r,V)= } \nonumber \\ 
&{}{}& S^{bulk}_{ref}(T_r,V)+\frac{1}{T} 
\left[ U(T_r,V) - U_{bulk}(T_r,V) \right] 
\label{eq:6c}
\end{eqnarray}
$S_{bulk}(T_r,V)$ is obtained as follows:
\begin{eqnarray}
\lefteqn{S^{bulk}_{ref}(T_r,V)=S^{bulk}_{ideal}(T_r,V)} \nonumber \\
&{}&+\frac{U_{bulk}(T_r,V)}{T_r}+ 
\int_\infty^V \frac{P^{bulk}_{exc}}{T_r} dV'
\label{eq:6d}
\end{eqnarray}
where $P^{bulk}_{exc}$ is the excess pressure of the bulk
and $S^{bulk}_{ideal}(T_r,V)$ is the entropy of the
ideal two component gas.

\begin{figure}
\includegraphics[width=80 mm]{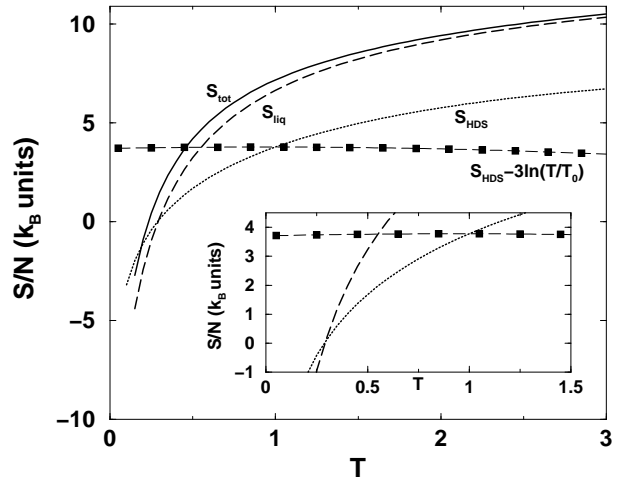}  
\caption{Total entropy of the confined system as obtained from
the thermodynamical integration (full line) as function of $T$.
The dashed curve is the entropy $S_{liq}$ obtained after subtracting
the contribution due to $W_{confin}$, see Eq.~\ref{eq:4b}.
In the figure are also reported the vibrational entropy $S_{HDS}$
(dotted line) calculated in the harmonic approximation and the
quantity $S_{HDS}-3ln(T/T_0$ (broken line with filled squares)
fitted with the polynomial $3.704+0.1555T-0.0835T^2$.
In the inset $S_{liq}$ and $S_{HDS}$ are reported on a larger
scale to better show the crossing point corresponding to
the Kauzmann temperature $T_K$.
All entropies are per particle and in units of $k_B$.
}
\label{fig:4}
\end{figure}

The result is the topmost curve shown in Fig.~\ref{fig:4}. 
Below the lowest investigated temperature $T=0.38$
the curve is extrapolated by an accurate polynomial fit.

At variance with the bulk in our case the effective density of the 
confined liquid is not constant, 
since the free volume accessible to the A and B particles
changes with the temperature due 
to the soft spheres interaction potential~\cite{pellarin1,pellarin2,attili1}.
The calculation of $S_{tot}$ has been performed along an isochoric
path and the internal energy of the confined liquid used in Eq.~\ref{eq:4}
contains also a contribution
$W_{confin}(T)$ due to the work done 
to change the effective density of the liquid  
inside the simulation box at constant volume. 
This contribution has to be subtracted to extract the entropy
of the liquid from which the 
the configurational entropy can be obtained
\begin{equation}
\label{eq:4b}
S_{liq}(T,V) = S_{tot}(T,V)-\int_{T_r}^T \frac{1}{T'} 
\left( \frac{\partial W_{confin}}{\partial T'} \right)_V dT'
\end{equation}
This integral can be calculated referring to a
corresponding bulk system simulated at the same pressures
and temperatures of the confined mixture, by considering
\begin{equation}
\label{eq:4c}
\frac{\partial W_{confin}}{\partial T} = 
\frac{\partial W_{confin}}{\partial V_{liq}} 
\frac{\partial V_{liq}}{\partial T}
\end{equation}
where $V_{liq}$ is the effective volume of the confined liquid.
$V_{liq}(T)$ can be derived by comparison with
an equivalent bulk at the same pressure.
The thermodynamical path followed by the equivalent bulk
and the corresponding densities will be published
in a separate paper~\cite{attili1}.
The result for $S_{liq}$ is also reported in Fig.~\ref{fig:4}.
We note that $S_{liq}$ of the confined system when compared
to that of the bulk~\cite{sciortino,coluzzi} assumes
higher values for high temperatures but approaches
the zero value at approximatively the same temperature as
the bulk. Therefore the enhancement of entropy due
to the additional disorder induced by the presence of the 
soft spheres seems to become less marked as the temperature is decreased.
 
Assuming the harmonic approximation to be valid also for our confined LJBM, 
from the eigenfrequencies obtained from the IS we can evaluate
the entropy of the harmonic disordered solid (HDS) with the formula
\begin{equation}
\label{eq:7}
 S_{HDS}= \sum_{i=1}^{3N-3} [ 1 - 
             ln \left( \beta \hbar \omega_i
                    \right) ]
\end{equation} 

The result is also reported in Fig.~\ref{fig:4}. 
In the same figure it is shown the quantity $S_{HDS}/Nk_B-3ln(T/T_0)$,
where $T_0=1$ has been chosen as
a reference temperature. It has a very weak T-dependence 
well fitted by a quadratic polynomial. This shows 
that almost all the $T$ dependence of the entropy
$S_{HDS}$  is contained in the term independent of the
frequency distribution. 

With the assumptions done for the vibrational spectrum
of the basins we can identify the $S_{HDS}$ with the
entropy of the disordered solid $S_{DS}$. In this way
the configurational entropy $S_{conf}$ is obtained from
the difference
\begin{equation}
\label{eq:8} 
S_{conf} \approx S_{liq}-S_{HDS}
\end{equation}
From Fig.~\ref{fig:4}, as best evident
in the blow up, we find that $S_{conf}=0$ at the
temperature $T=0.292 \pm 0.02$ which can be identified as
the Kauzmann temperature of the confined LJBM.

\section{Discussion and conclusions}

We have shown that for a LJBM 
the IS analysis can be performed also in confinement.
The absolute value of the
entropy of the confined liquid can
be obtained by thermodynamical integration by means
of a procedure where one refers to an equivalent bulk
system at the same temperature and volume as the confined liquid
for including the ideal terms. The result
must be corrected for the work done to change the
density of the confined liquid keeping constant the volume
of the simulation cell.
The correction is calculated by
comparison with a bulk liquid at the same pressure as the
confined one.
The combination of the IS analysis
and the thermodynamical integration technique allows to determine
the Kauzmann temperature $T_K$ of the system defined as the temperature
at which the configurational entropy vanishes.  

With the entropy of the disordered solid 
calculated in the  harmonic approximation as in the bulk,
we found that $T_K=0.292$ 
for the confined system to be compared with $T_K=0.297$
for the bulk. We observe therefore only a slight
decrease of $T_K$ upon confinement
while a much more marked decrease is instead detected
for the MCT cross-over temperature $T_C$.
We obtained in fact $T_C=0.356$ in confinement against
$T_C=0.435$ in the bulk~\cite{pellarin2,attili1}.

These results seem to confirm the connection between
dynamics and the thermodynamics energy landscape sampling
as a function of temperature~\cite{deben+still}. 
In the region close to $T_C$ the system is still
at relatively high temperature. The ergodicity is
assured by structural relaxations that require
cooperative rearrangement of large portions
of the liquid. This corresponds in the PEL
picture to a system that has sufficient kinetic energy
to sample a large portion of the PEL.
In this region the modification to the PEL induced
by the presence of the soft sphere matrix exerts
a strong influence on the particle motions
modifying substantially not only the $T_C$ but also
the critical exponents of the theory~\cite{pellarin1,pellarin2}.
On approaching the Kauzmann temperature the system
becomes trapped in a single minimum. 
In this situation only a small fraction of particles explores 
the configuration regions occupied by the soft spheres potential.

The confining matrix used in the present simulation
mimics the connected pore structure of systems with
high porosity like silica xerogels. It appears that the
confinement in this kind of system  does not shift
the thermodynamical liquid-glass transition 
but changes the way in which the configurational entropy
approaches the limiting Kauzmann temperature.
Further investigations will be necessary to understand:
(i) if the different behavior of $S_{conf}$ implies modifications
of the PEL, (ii)  if and how changes of porosity and/or size of the
confining spheres could modify the Kauzmann temperature.

\acknowledgments

We thank P. G. Debenedetti for valuable discussions.

\end{document}